\documentclass[manuscript]{aastex}

\usepackage{graphicx}

\newcommand{\degree}{\ensuremath{^\circ}}

\shorttitle{Weigelt knot ionization} 
\shortauthors{Remmen, Davidson, \& Mehner}  

\begin{document}

\title{Inverted Ionization Zones in Eta Carinae's ``Weigelt Knots''\altaffilmark{1}}

\author{ Grant N. Remmen\altaffilmark{2,3}, 
         Kris Davidson\altaffilmark{2}, 
         Andrea Mehner\altaffilmark{4} }

  \altaffiltext{1} {Based on observations made with the NASA/ESA {\it Hubble 
Space Telescope.\/} STScI is operated by the association of Universities 
for Research in Astronomy, Inc. under the NASA contract  NAS 5-26555.} 
  \altaffiltext{2} {School of Physics and Astronomy, University of Minnesota, 
       Minneapolis, MN 55455}   
  \altaffiltext{3} {California Institute of Technology, Pasadena, CA 91125}  
  \altaffiltext{4} {ESO, Alonso de Cordova 3107, Santiago de Chile}   

\email{}

\begin{abstract}
  
  The Weigelt knots, dense slow-moving ejecta near $\eta$ Car, are 
  mysterious in structure as well as in origin.  Using a special 
  set of spectrograms obtained with the {\it HST\/}/STIS, we have partially 
  resolved the ionization zones of one knot to an accuracy of about 
  10 mas.  Contrary to simple models, higher ionization levels occur 
  on the {\it outer\/} side of the knot, i.e., farther from 
  the star.  They cannot represent a bow shock, and no truly satisfying 
  explanation is yet available -- though we sketch one tentative 
  possibility.  We emphasize that STIS spectrograms provide far 
  more reliable spatial measurements of the Weigelt Knots than 
  {\it HST\/} images do, and this technique can also be applied to the knots' 
  proper motion problem.   

\end{abstract}


\section{The Weigelt Knots}    %%% SECTION 1  ===-=== 
   \label{sec:intro}

Many years ago, speckle imaging techniques revealed compact brightness 
peaks within 0.3{\arcsec} of $\eta$ Carinae \citep{weig86,hofm88}.   
Observations with the {\it Hubble Space Telescope (HST)} later showed 
emission-line spectra there \citep{davi95,davi97,weig12,hama12}.

Often called the ``Weigelt knots'' or ``Weigelt blobs,'' these objects 
have extraordinary attributes:  
(1) They move outward from the star at much lower speeds than $\eta$ Car's 
other ejecta, $V  \sim  40$ km s$^{-1}$ instead of $V  >  300$ km s$^{-1}$; 
(2) they were ejected significantly later than the star's Great Eruption 
of 1830--1860 \citep{weig95,davi97,dorl04,smit04};  and  
(3) they produce thousands of narrow emission 
lines of Fe$^+$, Fe$^{++}$, and other species \citep{zeth01,zeth12}.  
They are rather dense by nebular standards, $n_H \, >  \, 10^7$ cm$^{-3}$. 
Doppler shifts and astrometry indicate locations 300--1000 AU from the 
central star, and fairly close to the equatorial plane of the bipolar 
Homunculus ejecta-nebula.  Most authors suspect an origin in the 
``second eruption'' observed around 1890 (e.g.\ 
\citealt{weig95,davi97,smit04,smit12,weig12});  but, for reasons noted 
in {\S}\ref{sec:method} below, this surmise is difficult to prove.  
Indeed, since infrared images do not closely match visual-wavelength 
data \citep{arti11}, the features may conceivably be illusions caused 
by local minima in the circumstellar extinction, rather than physical  
condensations.  In summary, the Weigelt knots are known only in a 
rudimentary sense, and they have certainly not been explained.

Their emission lines are presumably excited, directly or indirectly, 
by radiation from the central binary star.  Observed spectra appear 
consistent with this hypothesis \citep{hama12}, and alternatives 
such as shock excitation have serious energy-supply difficulties 
({\S}\ref{sec:discuss} below).  Therefore the knots' spectra and 
ionization structure contain valuable information about the UV output 
of both central stars, if we can understand and quantify the morphology.
In this paper we report the first observations 
of spatial ionization structure in a Weigelt knot.

The \ion{Fe}{2}, [\ion{Fe}{2}], and other low-ionization features 
result from a combination of UV fluorescence plus ordinary thermal 
collisions in H$^0$/H$^+$, He$^0$ zones \citep{hama12}.   The \ion{He}{1}, 
[\ion{Ne}{3}], and [\ion{Ar}{3}] lines, however, arise in He$^+$ zones 
which require helium-ionizing photons, $h{\nu} > 25$ eV.  Since the 
primary star is too cool, these are thought to be supplied by the hot 
secondary star, see \citet{mehn10} and refs.\ therein.  Based on 
likely parameters, one expects high-ionization zones of 
He$^+$, Ne$^{++}$, etc., to exist in parts of the Weigelt knots that 
face toward the central star \citep{oste06,davi79}.  In other words, 
the simplest model predicts an inverse correlation between ionization 
level and distance from the star.  The size scale of the zones should 
help to constrain the local density values and the FUV output of the 
secondary star.

Here we describe spatially resolved measurements of ionization zones 
in Weigelt knot `C'.  But our main result is counter-intuitive, 
almost paradoxical:   {\it the stratification appears to be inverted,\/} 
with higher ionization at larger projected distances from the central 
star.  No satisfying explanation has yet been proposed.

                      %%%%  ===-===  %%%%  

\section{Observational Difficulties, and a Method based on Spectrograms}  
   \label{sec:method}   %%%  SECTION 2    

Three main unsettled observational problems require spatial resolution 
of the Weigelt knots:  their proper motions, sizes, and ionization 
structure.   Standard {\it HST\/} imaging (see, e.g., \citealt{dorl04}, 
\citealt{smit04}, \citealt{weig12}, and refs.\ therein) has proven 
inadequate for several reasons.   
  \begin{enumerate}  
  \item{  %% #1
    The image of the central star seriously contaminates those of the knots.
    With most available filters the peak brightness of each knot is less 
    than 4\% that of the central star, and the instrumental point spread 
    function (psf) has intricate structure at comparable levels out to  
    $r \, \sim \, 0.3\arcsec$.\footnote{  
        %%% FOOTNOTE    }    
        The one relevant exception is narrow-band filter F631N used with 
        the Wide Field Camera ({\it HST\/}/WFPC2).  In an image made with 
        this  filter, 10--20\% of the Weigelt-knot signal may represent 
        [\ion{S}{3}] $\lambda$6314 \citep{mehn10}.  But most of the 
        signal is due to other emission and/or reflection, and images give 
        no information about the relative contributions.   Moreover, the 
        {\it HST\/} psf is relatively broad at $\lambda > 6000$ {\AA}. }  
        An example of the small knot/star brightness ratio will be 
	noted at the end of {\S}3.3 below.  
      }   %%% (end of item 1)     
  \item{  %% #2       
    Standard processing techniques do not fully remove the 
    central star from {\it HST\/} images.  
    Experience shows that simple linear subtraction leaves illusory 
    features around $r \, \sim \, 0.2${\arcsec}, weak but sufficient 
    to perturb any measurement of the Weigelt knots.  Worse, the same 
    is true for standard deconvolution procedures.  The reasons 
    are beyond the scope of this paper, but concrete examples 
    can be seen in Figs.\ 1--3 of \citet{smit04}.  Those authors 
    attempted to deconvolve {\it HST\/}/ACS (Advanced Camera for 
    Surveys) images of $\eta$ Car, but their results show obvious 
    remnants of a circular ``ring of beads'' which is part of the  
    {\it HST's\/} basic psf (\citealt{kris11} and refs.\ therein). 
    In later ACS images with different {\it HST\/} roll angles, 
    some of those spots rotated with the instrument.\footnote{ 
       %%% FOOTNOTE  
       J.\ Ely 2010, priv.\ comm.;  unpublished data analyses.}     
    }  
  \item{  %%  #3 
    Every {\it HST\/} image of the Weigelt knots 
    samples a mixture of continuum plus emission lines.  Since the 
    continuum largely represents dust-reflected light from the star, 
    its spatial distribution very likely differs from that of the 
    line-emitting gas \citep{hama12,arti11}.  In order to estimate 
    the relative contributions, one must examine spectra 
    as discussed below. }  
  \end{enumerate} 
These circumstances cast serious doubt on any measurements of the knots 
in {\it HST\/} images.  A few special near-IR images are better (see, 
e.g.,  Fig.\ 4 in \citealt{arti11}), but there are not enough of them 
to give much information about ionization structure, motions, etc.

The difficulties become far less serious if we employ {\it HST\/} 
slit spectrograms rather than images.  Many of the Weigelt knots' 
emission lines have peak brightnesses far above the continuum.       
Moreover, we can take advantage of the narrowness of these lines 
(${\Delta}V \, <  \, 60$ km s$^{-1}$) to measure  
and remove contamination by the underlying star image.  The Space  
Telescope Imaging Spectrograph (STIS) is well adapted to this task. 

%%%%%%%%% ===-=== Figure 1  
\begin{figure}  
    \epsscale{0.7}  
\plotone{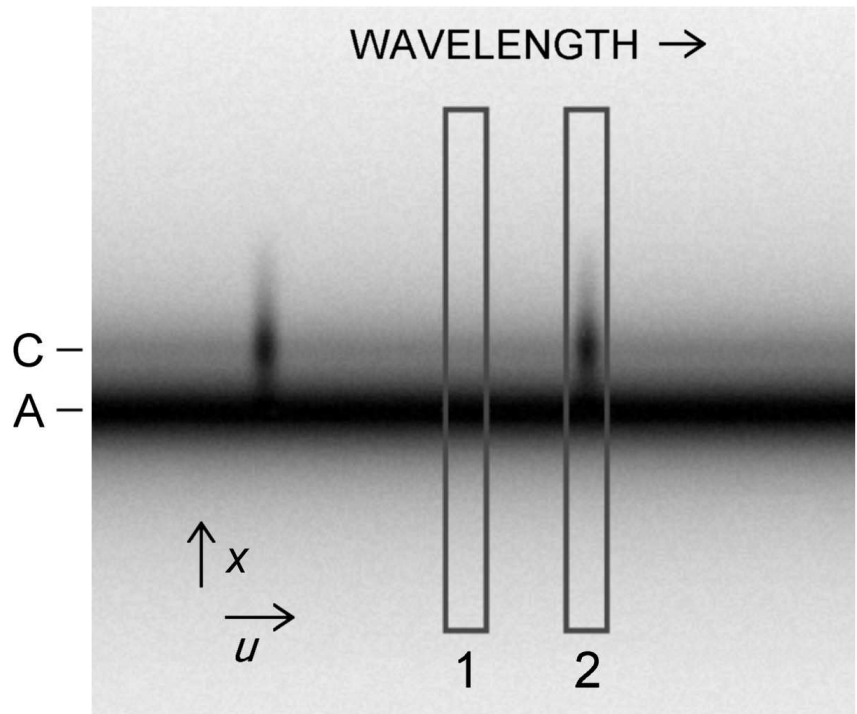} 
\caption{Narrow wavelength interval in a simulated spectrogram, showing 
   our basic measurement procedure.  `A' and `C' are the spectra of the 
   central star and of Weigelt knot C;  two narrow emission lines of the 
   latter can be seen.  Various complications are noted in the text.  }   
\label{fig:method} 
\end{figure}  
%%%%%%%%

Apart from complications noted in {\S}\ref{sec:data}, our basic 
approach is fairly obvious.  Figure \ref{fig:method} shows a small 
wavelength interval in a fictitious, idealized slit spectrogram.   
Light source A is the central star and C is one of the Weigelt knots.  
For simplicity, the star's spectrum is depicted as a continuum. 
Two emission lines of C can be seen in the figure, and here we measure 
the one on the right.  If $x$ denotes spatial position along the slit, 
we extract two separate spatial samples $f(x)$:  sample 1 avoids 
spectral features in both A and C, while sample 2 includes the chosen 
emission line in C.  Each of these can be a STIS/CCD column or a sum 
of adjoining columns.  After sample 1 is renormalized to correct 
for the wavelength dependence of the continuum flux, the difference 
$f_2(x) - f_1(x)$ represents only the emission line arising in knot C.   
If the spatial profile $f_1(x)$ depends on wavelength,  we can 
parametrize it by extracting more samples.

In practice this approach works quite well ({\S}\ref{ssec:removal}).  
The star subtraction is simpler and more robust than one can achieve 
in an {\it HST\/} image, because it requires only data from a small  
vicinity on the same spectrogram.  (In order to remove the star's 
optical halo from an ordinary image, either by subtraction or by 
deconvolution, one must derive the structure of its image based on optical 
modeling or another star image;  and $\eta$ Car's profile probably 
differs from a point-source psf.  The spectrogram method practically 
eliminates this difficulty.)   A mild non-linearity 
of the detector response would invalidate conventional 
methods of removing the central star optical halo in an image,  
but has only a second-order influence with our method.  And, 
most important, for many narrow emission lines the brightness ratio  
C/A is far greater than in any available non-spectroscopic image.

In order to obtain adequate spatial sampling with the STIS/CCD, 
one must employ ``dithered'' observations ({\S}\ref{sec:data}).  
Since no such data were obtained before 2010,  we have not yet attempted 
to apply our method to the proper motion and ejection-date 
puzzle.  Instead we focus on another, equally significant problem: 
the knots' ionization structure mentioned in {\S}\ref{sec:intro} above.

We chose narrow emission lines in four physically distinct categories:    
 \\ (A) Low ionization:  \ion{Fe}{2}, [\ion{Fe}{2}], and other species  
that can occur in either an H$^0$ or an H$^+$ zone.        
 \\ (B) Moderate ionization: [\ion{Fe}{3}].  Since the ionization 
      potential of Fe$^{+}$ is 16.2 eV, Fe$^{++}$ tends to coexist  
      with H$^+$ and He$^0$.  
 \\ (C) High ionization:  \ion{He}{1}, [\ion{Ar}{3}], and [\ion{Ne}{3}], 
      representing He$^+$ zones. (The \ion{He}{1} 
      lines are due to recombination.)  He$^0$, Ar$^+$, and Ne$^+$ 
      have ionization potentials of 24.6, 27.6, and 41.0 eV,  
      presumably requiring photons from the hot 
      secondary star.  See \citet{mehn10} and refs.\ therein.     
 \\ (D) Exotic:  \ion{Fe}{2} ${\lambda}{\lambda}$2508,2509. 
      Pumped by Ly$\alpha$, these extraordinary features have 
      laser-like properties almost unique in astrophysics  
      \citep{joha04,hama12,joha93,davi97}.

The high ionization lines were once suspected to come either 
from a diffuse region in which the Weigelt knots are imbedded, 
or perhaps from gas between the star and the knots \citep{vern05}.  
\citet{mehn10} showed, however, that 
[\ion{Ne}{3}] has brightness maxima near the knots.  Data 
available for that investigation could not resolve each knot.  
If the knots are real density concentrations, then the nature of 
photoionization leads one to predict stratified ionization zones 
as noted in {\S}\ref{sec:intro} above.   The relative 
location of \ion{Fe}{2} $\lambda\lambda$2508,2509 is potentially 
valuable because it is model-dependent:  it might occur in the 
moderate ionization zone because Ly$\alpha$ photons from the 
stellar wind have difficulty penetrating the low ionization zone, 
and some Fe$^+$ ions coexist with Fe$^{++}$.

\begin{figure}  
     \epsscale{0.7}  
 \plotone{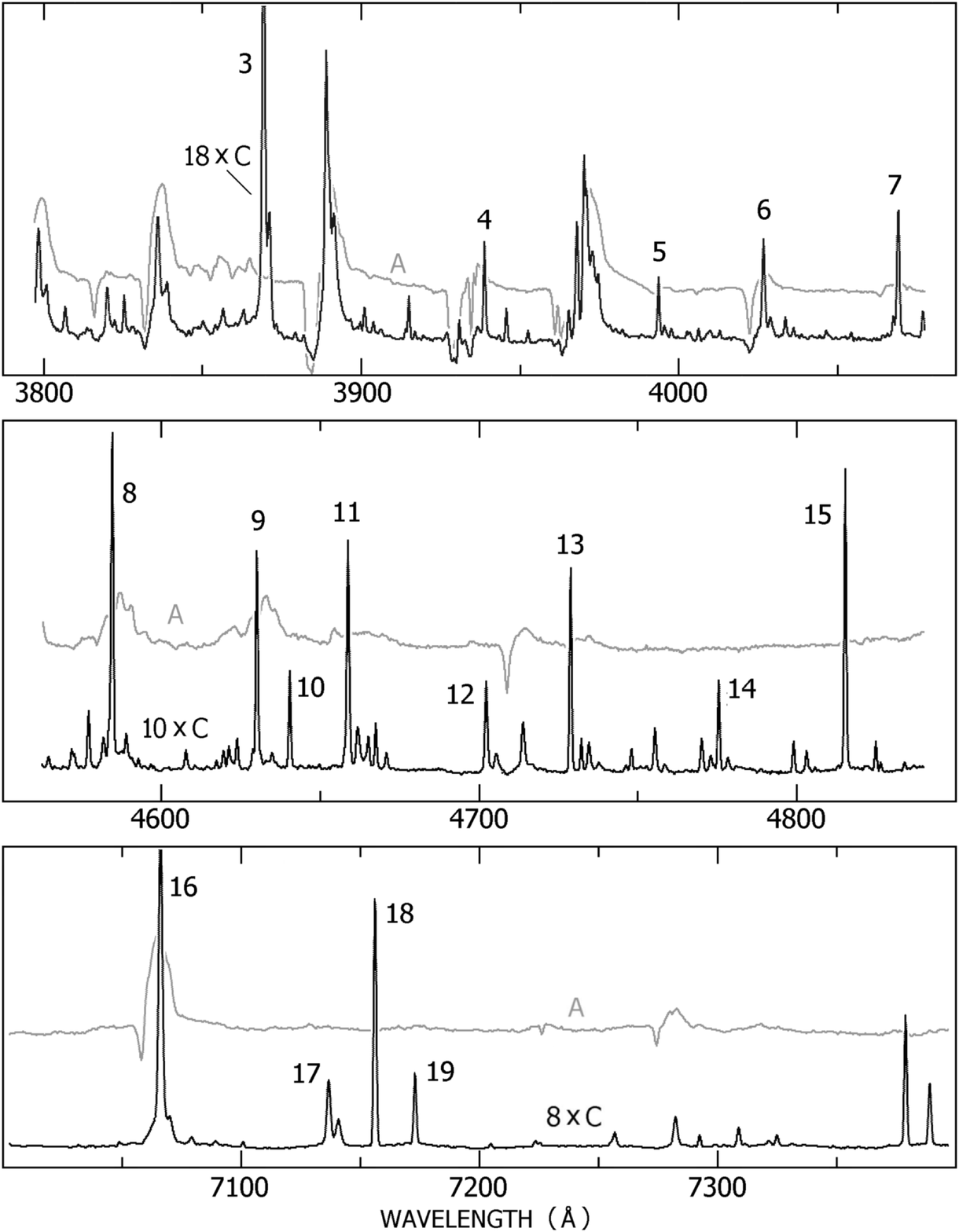} 
 \caption{HST/STIS spectra showing the emission lines listed in 
    Table \ref{tab:linelist}.  Gray tracings represent the central 
    star, while black tracings represent Knot C with magnified flux 
   scales.}     
 \label{fig:spectra}   
 \end{figure}

We measured the nineteen emission lines listed in Table \ref{tab:linelist}.  
Suitable data were available in four  wavelength intervals 
({\S}\ref{sec:data}), and we chose well-defined isolated lines, 
avoiding substantial features in the star's spectrum.  For instance, 
we omitted [\ion{Ne}{3}] $\lambda$3969 because it is confused with 
other strong features.   [\ion{Ar}{3}] $\lambda$7138 has too long a 
wavelength for {\it HST's} best resolution, but it gave  useful results 
({\S}\ref{sec:results}).  Spectral traces in Figure \ref{fig:spectra} 
show most of the selected emission lines.  The UV spectral region is 
omitted because $\lambda\lambda$2508,2509 hugely exceeds all 
other features there;   see, e.g., Fig.\ 8 in \citet{davi97}, 
Fig.\ 5.5 in  \citet{hama12}, and Fig.\ A.3 in \citet{zeth12}.    

                 %%%  ===-===   %%%    

\section{The Data Set and Detailed Methods}      %%  SECTION 3   
   \label{sec:data}

Since our procedure was carefully adapted to a non-routine purpose,
it requires a lengthy explanation. Readers interested mainly in the 
results, and willing to trust our precautions,  may choose to skip 
most of this section.

The STIS/CCD has a serious deficiency:  its 50.7 mas pixel size is too 
large to take good advantage of {\it HST's\/} spatial resolution 
in any single exposure \citep{davi06}.
Fortunately the spatial sampling can be improved by ``dithering,'' 
i.e., by taking separate observations at positions that differ by 
(integer + 0.5) pixels along the spectrograph slit.   Data of this type 
were obtained on 2010 March 3, as listed in Table \ref{tab:obslist}.  
The slit  was oriented along position angle 302{\degree} (close to NW by W), 
sampling the star and knot C \citep{weig12}.  Two nominal 
slit positions were used, with the same midline and differing in 
the observing plan by  
${\Delta}x \, \approx \, 228 \; \mathrm{mas} \, \approx \, 4.5$ pixels 
along the slit.  The 4.5-pixel difference, rather than 0.5, ensures 
that each $x$-locale is sampled by two different sets of physical pixels.  
We measured the true offsets ${\Delta}x$ as explained below.    
The available observing time allowed only enough wavelength coverage to 
include the best moderate and high excitation features plus 
[\ion{Fe}{2}] ${\lambda}{\lambda}$2508,2509 \citep{mehn10,hama12}.

Instead of the ``drizzling'' process usually applied to dithered 
{\it HST\/} images, we used a careful procedure 
described below.  Since it requires the original, geometrically 
unaltered CCD rows and columns, we worked with ``semi-raw'' data:   
flat-fielded, with cosmic ray hits and average underlying 
count levels removed, but omitting wavelength calibration and 
corrections for optical distortion.  
Cosmic ray removal was based on multiple exposures 
at each dither location  (``CR-SPLIT,'' $N$ in Table \ref{tab:obslist}).  
We considered treating each individual exposure separately and allowing 
for cosmic ray hits at a later stage in the process, but concluded 
that this would give little advantage in practice.  
In fact the final results were consistent enough to be 
self-verifying.

A few definitions are needed.   Let $u$ and $x$ denote CCD 
column and row number respectively, not necessarily integers because 
they may refer to an interpolated position.  To a first approximation,
$u$ represents wavelength while $x$ represents spatial position 
along the slit (Fig.\ \ref{fig:method}).  If $F(u,x)$ denotes 
intensity incident on the detector, then of  
course the CCD records only $f(u_m,x_n)$, the average of $F$ in each 
physical CCD pixel.  Regarding $f(u,x)$ as a continuous function, 
we estimated values between data points by 
cubic spline interpolation.  For the measurements in {\S}{\S}3.2--3.4 
below, we used a complete dither pair for each interpolation, so    
the $x$-interval between data points was approximately  
0.5 pixel rather than 1 pixel.\footnote{  
   %%%  FOOTNOTE
   Along any column in a single STIS spectrogram, spline 
   interpolation gives erratic results because of the inadequate 
   sampling \citep{davi06}.  Interpolation within a properly 
   dithered data set is far more satisfactory.  Strictly speaking, 
   our interpolation procedure included additional sub-steps that 
   turned out to be unnecessary. We omit them here because they had 
   no practical effect and explanations would be very lengthy. }

Deriving $F(u,x)$ from $f(u,x)$ is a non-trivial task, but fortunately 
the main results are apparent from the unenhanced $f$-profiles. 
Our procedure did {\it not\/} include deconvolution, mainly because 
no trustworthy psf was available and also because such a process 
may amplify the effects of pixel noise and other high-spatial-frequency 
defects. At any rate it proved to be unnecessary 
({\S}{\S}\ref{ssec:elmeasures} and \ref{ssec:other} below).

              %% ===-===  

\subsection{Measuring the spectral trace}   %%%  subsection 3.1  
   \label{ssec:trace}  

Spatial position is not exactly constant along a given CCD row;  
a spectrum ``trace'' -- the locus of a point-source continuum across 
a spectrogram -- is slightly tilted and curved.  Typically a STIS trace 
may shift by 1 row in about 150 columns ($dx/du \, \sim \, 0.007$), 
but this slope varies considerably among the various gratings and grating 
tilts.  Routine spectrum extractions take these facts into account, 
of course;  but our problem requires unusual spatial accuracy.  
We measured the central star's trace $x_{A}(u)$ in each spectrogram 
to high precision by the following method.

Consider an observed column vector $f(x)$ at a given $u$ in one 
exposure,  interpolated 
so that it is a continuous function.  It may be the average of several 
adjacent columns, avoiding emission features.  Local interpolation errors 
due to inadequate spatial sampling have very little effect when this 
procedure is completed.  The centroid of $f(x)$ is 
close to the star's position, but is mildly perturbed by the Weigelt 
knots, by noise, and especially by asymmetry of the psf.  A precise and 
consistent position measure can be obtained as follows.  First adopt 
a local weighting function 
${\phi}(s) = 1 - (s/a)^2$ for $|s| < a$ and 0 elsewhere, with 
$a =$ 4 pixels $\approx$ 203 mas.  At 
any  given position $x$ along the column, define a local quantity: 
  \begin{equation} 
     X(x) \; = \;  
        \frac{\int x' f(x') \, {\phi}(x' - x) \, dx'}
             {\int f(x') \, {\phi}(x' - x) \, dx'} \, .
  \label{eqn:locav}
  \end{equation}
Thus $X > x$ or $X < x$, respectively, on each side of the major peak 
of $f(x)$ due to the star.   A robust modified-centroid location $x_A$ is 
then defined by   
   \begin{equation}  
      X(x_A) \; = \; x_A \, .
   \label{eqn:centroid}
   \end{equation} 
Function ${\phi}(s)$ suppresses pixels that add noise but 
little information, and it also reduces pixelization effects.

For any given CCD column $u$, we search for the position $x_A$ that 
exactly satisfies (\ref{eqn:centroid}), and we adopt it as the position 
of the star in that column.  If an asymmetric psf causes $x_A$ to 
differ from the true position by a small amount, then other spatial 
features will have the same offset so their {\it relative\/} positions  
$x - x_A$ are meaningful to high accuracy.  We determine the trace 
$x_A(u)$ by fitting a cubic polynomial to the values measured in a set of 
well-spaced columns. A detailed analysis of this method would be 
too long for this paper, but the main advantages are:  
(1) it is conceptually simple, (2) the iterative 
procedure is easy to implement,  (3) it averages over pixel 
noise about as well as any method can,  (4) results are consistent 
without any need to know the parameters of the asymmetric STIS psf,  
and (5) if enough sample columns are used, the resulting $x_A(u)$ 
is quite insensitive to the STIS sampling problems described in 
\citet{davi06}.   This last fact is true because the slope of the 
trace $dx_A/du$ amounts to ``virtual dithering'' so far as 
the cubic fit is concerned.  In other words, $x_A$ coincides 
with a CCD row in some columns, it falls halfway between row 
midlines in some other columns, etc., and altogether these 
average out in the cubic fit.

In each spectrogram listed in Table \ref{tab:obslist}, we chose ten  
well-spaced column locations $u$, corresponding to wavelengths    
that avoided perceptible emission and absorption lines.   For each of 
these samples, the adopted $f(x)$ was the average of 5 adjoining CCD 
columns centered at $u$.  Then we used the ten sample values $x_A(u)$ 
to compute the least-squares cubic fit for $x_A(u)$ in that spectrogram.  
Based only on counting statistics, the formal error of each fit was less 
than 0.01 pixel or 0.5 mas across most of the observed range of $u$.   
(This statement is based on Monte Carlo simulations.)       
Systematic effects, e.g. due to the asymmetry of the STIS psf, can be 
larger but have almost no effect on the relative differences 
$x - x_A$ which ultimately determine our results.

Corresponding dither pairs (Table \ref{tab:obslist}) provide an obvious 
consistency test.  Ideally their traces $x_A(u)$ should differ 
by a constant ${\Delta}x$ = 4.5 pixels, the offset specified in the 
observing plan.     In fact, the r.m.s.\ value of 
$({\Delta}x - 4.5)$ for the six dither pairs in Table \ref{tab:obslist} 
is 0.009 pixel $\approx$ 0.5 mas.  Variations across the CCD are larger
because of image distortions in the STIS. 
An example:  Evaluating the cubic-fit dither offsets ${\Delta}x$ at 
wavelengths of relevant emission lines in the short-exposure 
4565--4845 {\AA} dither pair, we find 
4.484 $\lesssim {\Delta}x \lesssim 4.510$;  a  range of 
0.026 pixel $\approx$ 1.3 mas.  Altogether, the estimated star position 
$x_A(u)$ appears to be consistent within $\pm$ 1 mas at most wavelengths, 
only 2\% of the instrumental resolution.   Of course this high quality 
required a large number of data pixels for each fit.   
The instrumental variations of spatial  
scale are negligible for our purposes, partly because we employ 
only a few of the CCD rows running through the central part of the   
detector.  The main point is that errors in the trace $x_A(u)$ are 
much smaller than the effects of interest which exceed 10 mas 
({\S}\ref{sec:results} below).  Some extra tests not worth detailing 
here were also applied, such as comparisons between independent 
dither pairs.  They all had satisfactory outcomes.

Therefore, when examining the spatial position of an emission line in 
the Weigelt knot C, we can safely refer to a true spatial coordinate  
  \begin{equation}   
    z \; = \;  x - x_A(u)    
  \label{eqn:zdef}  
  \end{equation} 
where $x_A(u)$ is known to high accuracy.

\subsection{Subpixel modeling}   %%%  subsection 3.2  
   \label{ssec:subpix}  

As noted earlier, dithering along the slit is essential because the 
STIS/CCD spatial sampling is too sparse to take full advantage of 
{\it HST's\/} basic resolution.\footnote{ 
   %%%  FOOTNOTE
   This statement is not equivalent to the distinction between 
   $f(x)$ and $F(x)$ mentioned earlier, though it is related. 
   See \citet{davi06}. }   
For any given CCD column $u$, a dither pair of spectrograms provides 
two sample vectors:  
  \begin{equation}
    f_{n}^{(1)}  =  f(n)   \ \  \mathrm{and}    \ \
    f_{n}^{(2)}  =  f(n + {\Delta}x - 4) \, ,  
  \label{eqn:samplevectors}  
  \end{equation}
where the dither offset ${\Delta}x$ is practically a known constant. 
(Strictly speaking, it's a known function of $u$ based on the 
individual exposures' measured traces $x_A(u)$.) 
The two vectors together provide a sampling interval of about 
0.5 CCD-pixel or 25 mas,  nearly adequate for {\it HST's\/} resolution 
according to the Nyquist criterion.   For each relevant column or 
sum of adjoining columns, we used cubic spline interpolation to derive 
a continuous function $f(x)$.  (Of course this interpolation employed the true 
${\Delta}x$ values, not the uniform nominal spacing of 0.50 pixel.)   
Then we shifted each $f(x)$ to produce $f(z)$, a spatial 
distribution relative to the star's position (eqn.\ \ref{eqn:zdef}).

We half-expected the two parts of each dither pair to  differ perceptibly 
in their intensities and psf's -- due, e.g., to variations of the 
``jitter'' in the {\it HST\/} pointing, slight drifts  perpendicular to 
the slit, electronic subtleties, etc.   In fact no such differences 
were found.

\subsection{Removal of the central object}  %% subsection 3.3 
   \label{ssec:removal}  

As noted in {\S}\ref{sec:method} and Figure \ref{fig:method}, 
we must subtract a ``continuum'' spatial profile $f_1(z)$ from 
the spatial distribution $f_2(z)$ measured at each narrow 
emission line.    The underlying $f_1$ represents mainly the 
central star but it also includes continuum and dust-reflected 
light from knot C.  In order to estimate the relevant $f_1$'s,  
we sampled spatial profiles at various wavelengths that had no 
perceptible emission features.  In order to avoid biased sampling 
in the $x$-direction,  we included pairs of wavelengths whose 
trace positions $x_A(u)$ differed by substantially non-integer 
numbers of pixels ({\S}\ref{ssec:trace} above).

Within each observed wavelength interval, $f_1(z)$ varied with 
wavelength less than one might expect.  The optical diffraction limit 
by itself would imply a narrower spatial psf near the short-wavelength 
end of an interval.  But this is counteracted by poor STIS focussing 
on the shorter-wavelength side of the CCD, see \citet{davi06} and 
the instrument handbook.\footnote{ 
   %%% FOOTNOTE 
   http://www.stsci.edu/hst/stis/handbooks/.   } 
   %%%  
For instance, we found widths between 1.524 and 1.561 pixels (FWHM) 
for $f_1(z)$ across the interval 4585--4810 {\AA} -- a range of 
only 2.4\% even though the wavelength varied by 5\%.

 \begin{figure}  
   \epsscale{0.75}  
 \plotone{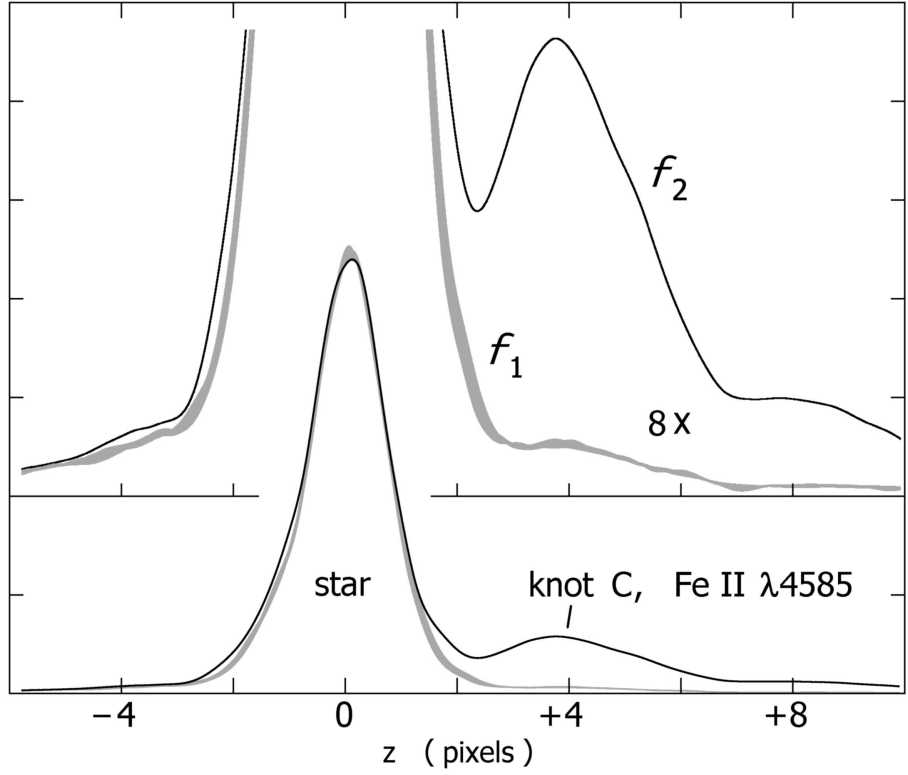} 
 \caption{Observed spatial profiles along the slit near 
    $\lambda \sim 4585$ {\AA}.  The two panels differ only in their 
    vertical scales.  In each panel, the upper trace $f_2$ represents 
    a wavelength at the center of the \ion{Fe}{2} $\lambda$4585  
    emission line.  The lower trace $f_1$ shows the {\it envelope\/} of 
    four independent samples at wavelengths with no perceptible 
    emission features.  The five samples were renormalized so they 
    approximately match in the brightest part of the star image. }  
 \label{fig:4585profs}  
 \end{figure}  

In each panel of Figure \ref{fig:4585profs}, the lower trace $f_1(z)$ 
depicts the envelope of 4 separate profiles in that 
wavelength range.  Larger variations were found in the other three 
wavelength ranges, but for each emission line we used nearby samples 
of $f_1(z)$.  In the important case of [\ion{Ne}{3}] $\lambda$3870, 
for example, we found a FWHM between 1.663 and 1.669 pixels 
across the interval 3850--3950 {\AA}.  These $f_1$ widths 
represent combinations of the basic instrumental psf, imperfect 
optical focus, {\it HST\/} jitter, and very likely a real non-point-like 
width of $\eta$ Car's wind;  for our purposes there is fortunately 
no need to know the relative size of each effect.  The star's 
spectrum is more complex at UV wavelengths 2480--2680 {\AA}, but 
the strange \ion{Fe}{2} $\lambda\lambda$2708,2709 
lines are extremely bright in the Weigelt knots, easy to 
separate from the star.

Subtracting $f_1(z)$ from each narrow-emission-line profile $f_2(z)$ 
therefore presented no serious difficulty.   
This is manifestly true for the brightest measured lines, which 
considerably exceeded the underlying $f_1(z)$.  Figure \ref{fig:4585profs} 
shows one such case, \ion{Fe}{2} $\lambda$4585 (see    
Table \ref{tab:linelist}).   For each narrow emission line we 
simply used the measured profile $f_1$ that was closest in wavelength.   
In each case $f_1$ was renormalized to match the integrated brightness  
of $f_2$ at the star's peak.   (Strictly speaking, we based the 
adjustment factor on the maximum values of 
    $q(z) \, = \, \int \psi(z-z') f(z') dz'$ 
for $f_1$ and $f_2$, where $\psi$ is a parabolic weighting function 
only 1 pixel wide.  Other definitions give practically 
the same results.)  
This renormalization is not exactly valid if the narrow emission 
line extends across the star image;   but any 
resulting error in the difference $f_2 - f_1$ is very small at the 
location of knot C, because $f_1(z_C) << f_1(0)$, see Fig.\ 
\ref{fig:4585profs}.  In recent years the narrow emission lines along 
our line of sight to the star have been quite faint compared to the 
star itself \citep{mehn10}.

 \begin{figure}  
    \epsscale{0.7}  
 \plotone{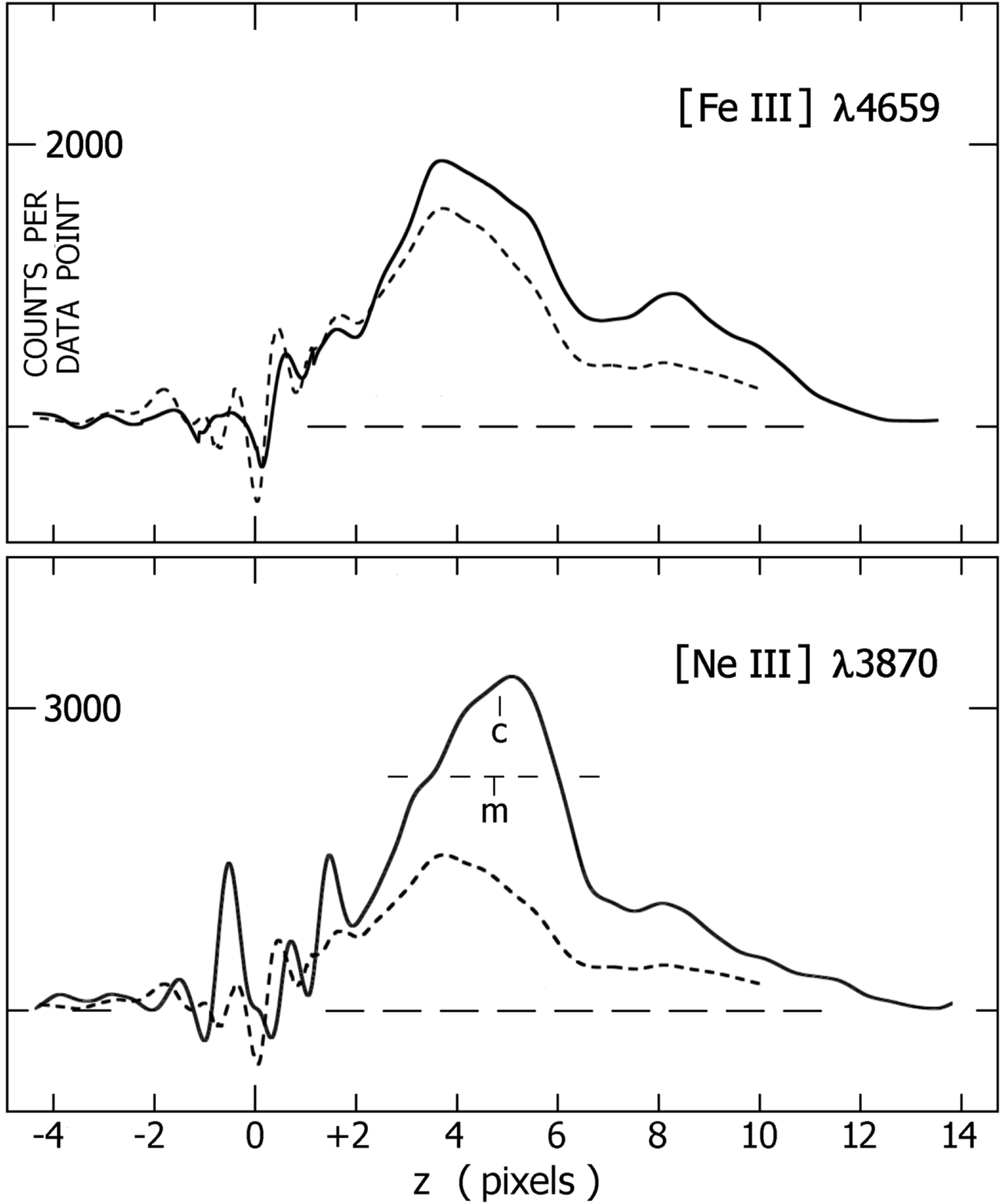} 
 \caption{Net spatial profiles $g(z)$ for the [\ion{Fe}{3}] $\lambda$4659 
    and [\ion{Ne}{3}] $\lambda$3870 emission lines in knot C.  For the 
    latter, points `m' and `c' are ``midpoint'' and ``centroid'' positions 
    defined in {\S}3.4.  Dashed curves show the average  spatial profile 
    of four [\ion{Fe}{2}] lines, see text {\S}\ref{sec:results}. }  
 \label{fig:netprofs1}   
 \end{figure}

 \begin{figure}  
     \epsscale{0.6}  
 \plotone{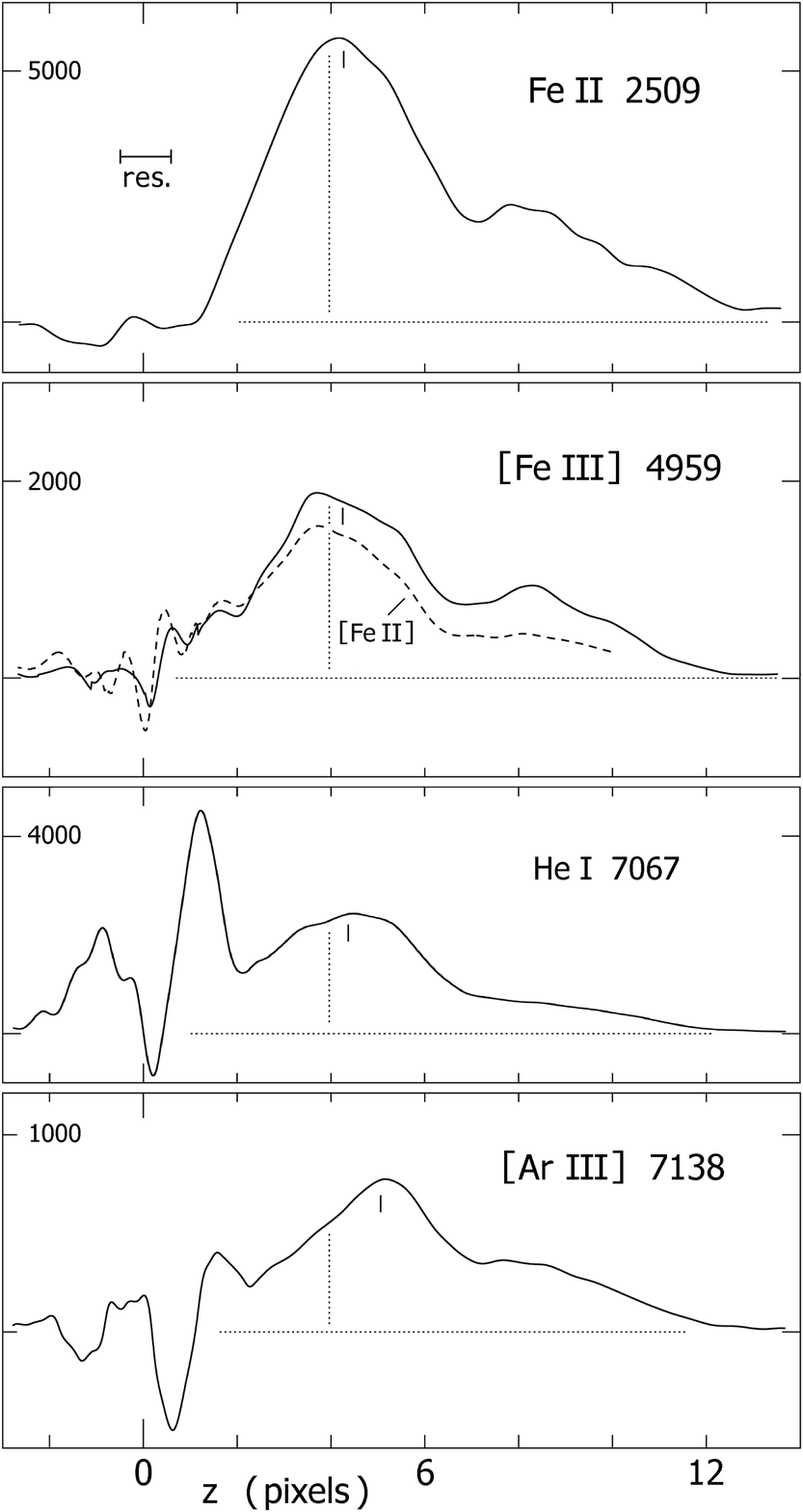} 
 \caption{Spatial profiles $g(z)$ for several emission features with 
  differing characteristics.  A small vertical mark shows $\hat{z}$ 
   for each feature, defined in {\S}3.4. The dashed curve is an 
   average of four [\ion{Fe}{2}] lines, the same as in 
   Fig.\ \ref{fig:netprofs1}. 
   The vertical dashed line shows their average $\hat{z}$. }   
 \label{fig:netprofs2}   
 \end{figure}

Figures \ref{fig:netprofs1} and \ref{fig:netprofs2} show examples of 
net profiles $g(z) = f_2(z) - f_1(z)$.  Numerical oscillations occur 
at the location of the star, $-2.3 \, \lesssim \, z \, \lesssim \, +2.3$, 
because the net values there are differences between two very large 
quantities which are nearly equal but slightly imperfect. 
As one expects  from Figure \ref{fig:4585profs}, the Weigelt knot is 
represented  quite well for $z \gtrsim 2.5$ pixels -- much better 
than in any non-spectroscopic {\it HST\/} image.  Separate dither pairs 
(independent sets of STIS exposures) gave $g(z)$ profiles that mutually 
agreed to within the uncertainties set by counting-noise.  \ion{He}{1} 
profiles tend to be less satisfying than the others, for a reason 
noted in {\S}\ref{sec:results} below.

Incidentally, Figure \ref{fig:4585profs} illustrates the difficulty 
of measuring a Weigelt knot in standard images.  In continuum light, 
knot C appears only as a small bump in the $f_1$ tracing near 
$z \sim 4$ pixels, scarcely brighter than the star's psf at that 
location.  By contrast, the narrow emission line greatly exceeds the 
star in that locale;  compare curve $f_2$ to $f_1$ in the figure.            

                %%  ===-===  

\subsection{Measuring the positions}   %%%  subsection 3.4  
   \label{ssec:elmeasures}  

The most reliable part of the net spatial profile $g(z) = f_2(z) - f_1(z)$ 
is obviously near its peak, see Figures \ref{fig:4585profs} and 
\ref{fig:netprofs1}.  Therefore we based our position measurements 
on the 70\%-of-peak level.  For each narrow emission line the procedure 
was as follows.   Begin with the available dithered data points along 
the appropriate CCD column, i.e., at half-pixel intervals of $z$. 
Via cubic splines, these define a continuous 
spatial profile $g(z)$.   Denote by $z_a$ and $z_b$ the two 
places where $g(z) = 0.7 \, g(\mathrm{peak})$.   Then, if 
$h(z)$ is the quantity $g(z) - 0.7 \, g(\mathrm{peak})$,  
one can easily calculate a ``midpoint'' and ``centroid'':   
   \begin{equation}  
        z_m \; = \; \frac{z_a + z_b}{2}   
        \ \ \ \mathrm{and}  \ \ 
        z_c \; = \; 
           \frac{\int z \, h(z) \, dz}{\int h(z) \, dz} \; ,  
   \label{eqn:zmzc} 
   \end{equation}  
with integration limits $z_a$ and $z_b$.  The difference $z_c - z_m$ 
indicates asymmetry near the line peak.  Examples are shown in 
the lower panel of Figure \ref{fig:netprofs1}.  In most cases we 
use the simple average $\hat{z} = 0.5 \, (z_m + z_c)$.

We estimated statistical uncertainties by performing random  
simulations with profiles like those shown in Figure \ref{fig:netprofs1}.  
Several types of statistical error occur.  
(1) Most important is the counting noise associated with the square root 
of $f_1 + f_2$.  For a net profile $g$ with a peak of  
1500 counts per data point, the resulting rms error in either $z_m$ 
or $z_c$ was found to be roughly $\pm$0.07 pixel.  Noise errors 
are of course worse for fainter profiles.  
(2) Imperfect spatial sampling also has an effect, because the precise 
cubic-spline fit $g(z)$ depends on the location of the pixel 
array relative to the spatial profile.  This 
depends on the profile shape, but we estimated typical rms 
errors in the range $\pm$0.003 to $\pm$0.01 pixel.  These do 
not depend on the strength of the emission line. 
(3) According to {\S}\ref{ssec:trace} above, errors in the spectral 
trace $x_A(u)$ are not worse than $\pm$0.03 pixel.  Uncertainties 
quoted above are formal statistical estimates, and {\it systematic\/} 
errors may be larger ({\S}\ref{ssec:stats1} below).

In principle, one-dimensional Lucy-Richardson deconvolution can enhance 
the spatial resolution; but a few trial examples showed no worthwhile 
improvement with these data.  Knot C appears to be 
partially resolved without deconvolution, having   
FWHM $\sim$ 3.6 pixels $\sim$ 180 mas (Fig.\ \ref{fig:netprofs1})
-- i.e., almost 3 times as wide as the overall psf.  Since deconvolution 
tends to magnify small numerical irregularities and noise at high 
spatial frequencies, we chose not to employ it here.

                %%%  ===-===    

\section{Results}   %%  SECTION 4
   \label{sec:results}

Figures \ref{fig:netprofs1} and \ref{fig:netprofs2} show some of the 
measured spatial profiles $g(z)$.  As a reference for comparison, 
in each figure a dashed curve shows the unweighted average of four 
[\ion{Fe}{2}] features labeled 10, 13, 14, 15 in  
Tables \ref{tab:linelist} and \ref{tab:measures}.    

   \begin{figure}  
     \epsscale{0.8}  
   \plotone{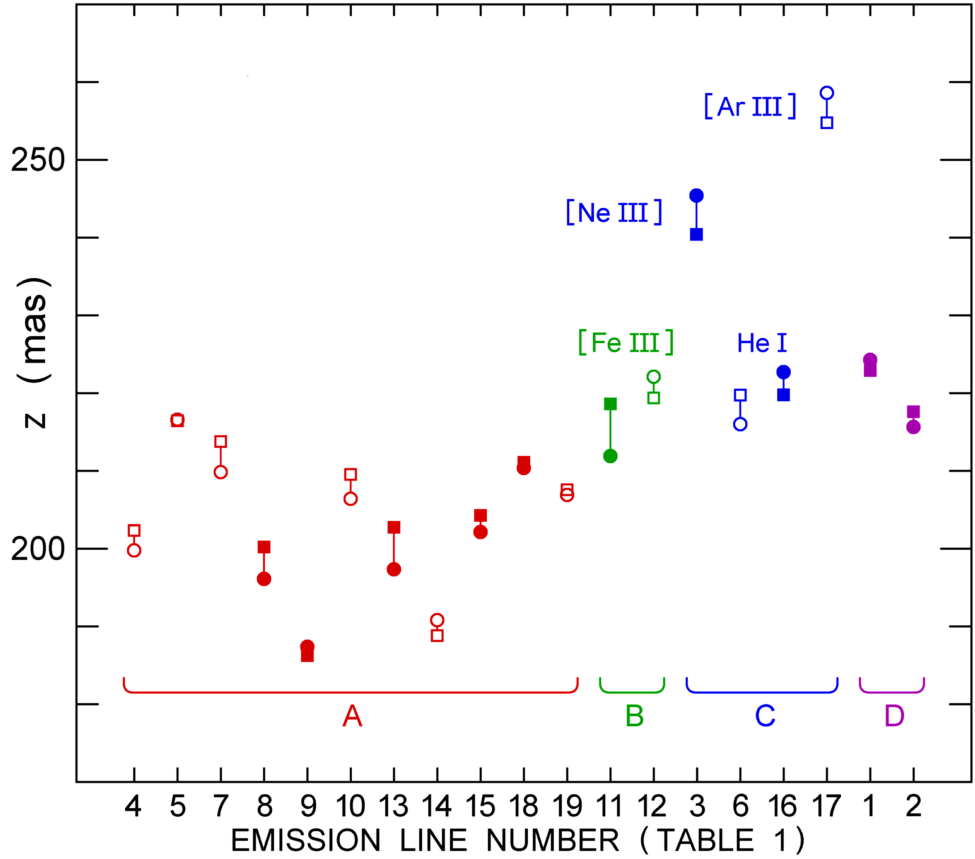} 
   \caption{Measured distances of emission peaks from the star, 
     sorted by ionization category.  These are the results listed 
     in Table \ref{tab:measures};  filled symbols represent 
     features with higher count rates.  Circles and squares 
     indicate $z_c$ and $z_m$ respectively, see text.}   
   \label{fig:zvals}   
   \end{figure}

The most surprising result concerns spatial location $z$ as a 
function of ionization level, shown in Tables \ref{tab:measures} and 
\ref{tab:zlist} and Figure \ref{fig:zvals}.  
{\it Highly excited features obviously tend to occur farther 
from the star, i.e.\ at larger $z$, not closer to the 
star as was expected\/} ({\S}\ref{sec:intro}).  Two very different 
statistical analyses in {\S}4.1 and {\S}4.2 below confirm the 
reality of this effect.

Before reviewing likely errors,  it is important to recall that our 
measurement procedure was blind to ionization and excitation level;  
the various emission lines occurred at essentially random columns 
on the detector, and they were all treated alike.   Note also that 
[\ion{Ne}{3}] and [\ion{Ar}{3}], the conspicuously highest points in 
Figure \ref{fig:zvals}, probably represent Category C better than the 
\ion{He}{1} lines do.  [\ion{Ne}{3}] and [\ion{Ar}{3}] originate 
mainly in the highest-temperature gas, while the \ion{He}{1} recombination 
lines are less sensitive to temperature and tend to favor  
the coolest parts of the He$^+$ zones \citep{oste06,davi79}. 
Moreover, $\eta$ Car's stellar wind produces substantial \ion{He}{1} 
features, which may perturb our \ion{He}{1} positional measurements toward 
slightly smaller values of $\hat{z}$;  indeed this is obvious for 
\ion{He}{1} $\lambda$7067 in Figure \ref{fig:netprofs2}.  Therefore 
Table \ref{tab:zlist} summarizes the $\hat{z}$ values for two versions of 
Category C:  with and without the \ion{He}{1} lines.  

      %%%  ===-===  

\subsection{Statistical significance based on measurement quality}     
   \label{ssec:stats1}  %% subsection 4.1  

One form of confidence level involves the   
measurement uncertainty.  We can estimate this quantity based on  
the scatter of either $z_m$ or $z_c$ among the nine  \ion{Fe}{2} 
and [\ion{Fe}{2}] features listed in Table \ref{tab:linelist}.  In 
principle the forbidden lines might differ from permitted lines 
because the latter have higher upper energy levels;  but in fact 
we find no statistically significant difference between the six 
[\ion{Fe}{2}] lines and the three \ion{Fe}{2} features.  If all 
of them have the same true average position, then the scatter in 
measured values indicates an r.m.s.\ error of about $\pm$0.17 pixel 
or $\pm$8.6 mas for both $z_m$ and $z_c$.  This is roughly twice as 
large as the semi-formal statistical uncertainty estimated  
in {\S}\ref{ssec:elmeasures}.  Therefore, undiagnosed effects of 
the order of $\pm$7 mas -- instrumental subtleties and/or imperfect 
assumptions about the emission -- probably dominate the error budget.  
This is a common, or even usual, circumstance for sensitive astronomical 
measurements.\footnote{
    %%% FOOTNOTE \
    Real positional differences may exist among the \ion{Fe}{2} and 
    [\ion{Fe}{2}], but theoretically there should be almost no 
    difference between lines 10 and 13 (multiplet 4F) or between 
    lines 14 and 15 (multiplet 20F). }  
    %%%
Such effects can probably be treated like random errors here, because, 
as noted above, the wavelengths, detector locations, and intensities 
were not seriously correlated with emission-line category or excitation.  
If all the $\hat{z}$ values are systematically too high or too low,  
this has no effect on differences between emission categories.

Now consider the unweighted average of $\hat{z}$ for each ionization 
category.  If the r.m.s.\ error 
for an individual line is $\pm$8.6 mas as suggested above, then the 
formal averages are 203 $\pm$ 2.5 mas for ionization category A 
($N = 11$ spectral features) 
and 229 $\pm$ 3.5 mas for categories B and C together ($N = 6$);  
a 6$\sigma$ difference.  If, pessimistically but somewhat illogically, 
we base the individual measurement uncertainty on the larger scatter 
within data set B$\, \cup \,$C, then the difference is still 
3.4$\sigma$.     If we omit the weakest lines in each 
category, or if we omit the intermediate category B, then the confidence 
level becomes stronger.  In summary, the conclusion that 
$\hat{z}(\mathrm{B \, \cup \, C}) > \hat{z}(\mathrm{A})$ 
is quite well established so far as random errors are concerned.

The relatively large scatter in $\hat{z}$ within categories B and C is 
not surprising.  [\ion{Fe}{3}] represents a lower-ionization 
zone than [\ion{Ne}{3}], [\ion{Ar}{3}], and \ion{He}{1}, while all 
these lines span a range of temperature dependences.  
 
           %%%  ===-===  

\subsection{A different approach to statistical significance}   
    %% subsection 4.2 
   \label{ssec:stats2}

If one is skeptical about the r.m.s.\ measurement errors reported  
above, another form of reasoning does not require them.   
Consider, for example, the following statement about the strongest  
emission lines, the filled symbols in Fig.\ \ref{fig:zvals}.  
{\it All three of the $\hat{z}$ values in ionization categories 
B and C exceed all five of those in category A.} 
If they were all random samples of one 
population, the probability of this outcome would be less than 0.02.  
This type of test is valid for any reasonable population distribution. 
The fainter lines strengthen the case.   
Suppose that the eleven measurements in category A and the six 
in categories B and C  constitute two sets of random samples. 
(Differences in quality may alter this assumption,   
but more elaborate analyses lead to the same conclusion.)   
{\it All of the $\hat{z}$ values in set $\mathrm{B} \, \cup \, \mathrm{C}$  
are higher than the second-largest value in set} A.  If both sets 
were drawn from one population, the probability of this outcome 
would be less than 0.0006.  If we further note that only one value 
in set B $\cup$ C lies below the highest in set A, this becomes 
a problem in multinomial coefficients (e.g., \citealt{degr02}) 
and the probability falls below 0.0002.  Set B $\cup$ C thus differs 
from set A with a high confidence level, whether we use 
all the spectral features or only the best ones.   This statement 
does not require any knowledge of the measurement errors. 
   
             %%%  ===-===  

\subsection{Concerning systematic errors}   %% subsec 4.3 
  \label{ssec:systematics}

Although the measurement procedure was blind to ionization state 
as noted above, there is an obvious danger in the small number of suitable 
high-ionization features.  Subtle effects involving location on the 
CCD detector might conceivably influence the results merely because 
the distribution  
of [\ion{Ne}{3}], [\ion{Ar}{3}], and [\ion{Fe}{3}] lines was sparse. 
Fortunately we can assess this possibility via the low-ionization lines.  
In the best grating-tilt wavelength interval, 4565--4845 {\AA} 
(Table \ref{tab:obslist}), three category A lines have smaller 
wavelengths than [\ion{Fe}{3}] and three have larger wavelengths;  together 
they show no trend large enough to affect the result for [\ion{Fe}{3}].  
The 3795--4075 {\AA} interval is less satisfactory because all three 
measureable category A lines there have longer wavelengths than 
[\ion{Ne}{3}] $\lambda$3870 -- but the $\hat{z}$ value for that feature 
is so large that a hypothetical wavelength-dependent effect would 
need to be far greater than any comparison lines suggest.   A similar 
remark applies to [\ion{Ar}{3}].   
In summary:  The low-ionization features appear to confirm what 
one expects from the nature of the instrument, i.e., that there is no 
serious wavelength-dependent error in our measurement procedure 
within each spectrogram.\footnote{
    %%% FOOTNOTE
    Optical distortions are negligible in the narrow range of 
    CCD rows used here, see the instrument handbook at 
    www.stsci.edu/hst/stis/handbooks/.  In order to nullify 
    the [\ion{Ne}{3}] result, one would need a spatial-scale 
    variation of the order of 10\% between 3870 {\AA} and 4000 {\AA}. }   

                %%%  ===-===  

\subsection{Other potentially important findings}    %%% subsec 4.4 
   \label{ssec:other}   

 Weigelt knot C is partially resolved in the light of emission lines,  
with FWHM $\sim$ 180 mas compared to about 80 mas for the star image.  
Therefore the true, deconvolved FWHM is probably between 150 and 170 
mas, or about 370 AU at $\eta$ Car's distance.   Near-IR data give 
the same extent  for dust in the knot,  see Figure 4 in \citet{arti11}. 
Ionization categories A, B, C, and D spatially overlap each other 
to a great extent.  This is not surprising, since we view the 
ionization structure from an oblique angle relative to the 
star-knot radial vector.  Note, however, that the spatial profiles 
for [\ion{Ne}{3}] and [\ion{Ar}{3}] are strongly skewed toward 
larger $z$.  This fact by itself is enough to distinguish those 
two highest-ionization features from most of the others.  
On average the low-excitation emission appears skewed in the 
opposite direction.  
We have not attempted to model the asymmetric spatial profiles 
shown in Figures \ref{fig:netprofs1} and \ref{fig:netprofs2}.

Those figures also show a second obvious brightness peak around 
$z \approx$ 8.4 pixels $\approx$ 430 mas.   Based on this and 
previous hints (e.g., {\S}3.2 in \citealt{mehn10}), evidently  
the emission morphology has numerous local spots, not just the 
three classic Weigelt knots.   The example shown in our data must 
be roughly aligned with the star and Knot C.  Faint material 
can be perceived near that location in some {\it HST\/} images 
\citep{smit04}, but it seems uncertain there for the reasons 
noted in  ({\S}\ref{sec:method}) above.  In any case the 
STIS results define this outer knot fairly well, and they show 
that it is almost half as bright as Knot C in 
emission-line categories B, C, and D\/ (Figs.\ \ref{fig:netprofs1} 
and \ref{fig:netprofs2}).   Since the outer knot does not show 
a clear peak in the lower ionization features, we did not attempt 
to detect ionization structure there.   The most effective way to map 
the region is to use STIS spectra with spatial dithering parallel 
to the slit -- another task that no one has attempted.

Small Doppler velocity effects may accompany the 
positional differences shown in Fig.\ \ref{fig:zvals}. 
Measurements using  ordinary methods show no significant velocity 
trends among the  categories of emission lines, but the 
differences may be too small to detect routinely.  The individual 
lines' velocity profiles probably differ as well.  A special velocity 
investigation with unusually precise wavelength calibrations, etc., 
is beyond the scope of this paper, because it would require as much 
additional effort as the steps reported above.  \citet{zeth01} and 
\citet{zeth12} list many emission velocities in the Weigelt knots 
with ordinary accuracy, including all the features in 
Table \ref{tab:linelist}.

      %%%  ===-===  

\section{Discussion}   %%% section 5  
   \label{sec:discuss}

The unexpectedly inverted excitation structure -- with higher-ionization 
species relatively farther from the star rather than closer --   
is not easy to explain.  Very likely it is a clue to some previously 
unrecognized aspect of the Weigelt knots' morphology.

The most obvious idea, excitation by an outer bow shock,  
fails for at least two reasons.   First, this conjecture would 
require ambient gas moving either {\it slower\/} than Knot C relative 
to the star, or else inward.  Since the knots' velocities are themselves  
very slow in the context of $\eta$ Car, and the vicinity should have 
been swept out by two eruptions and the stellar wind, this seems 
very unlikely.  There is no emission-line evidence 
for motionless or inward-moving gas.  A second objection concerns the 
energy budget.  The mass of visible, unobscured material in Weigelt 
knots B, C, and D is of the order of 0.01 $M_\odot$ \citep{hama12,davi95},
implying kinetic energy $\sim \ 10^{44}$ ergs at the observed 
speed of roughly 40 km s$^{-1}$.   Even if this is converted into 
emission lines within only 100 years, the total 
luminosity of all the lines would be less than $10^{34.6}$ ergs s$^{-1}$, 
several orders of magnitude too small.  Moreover, this is an optimistic 
estimate since the knots do not appear to be 
decelerating that rapidly, and because most of the energy of a 
shock dissipates through expansion rather than line emission. 
The bow-shock idea therefore appears highly implausible.

Lesser shocks or other material waves moving through the knots are 
similarly unappealing.  In order to excite even a fraction of the 
observed  emission, they would need to carry so much energy and 
momentum that the knots would be disrupted within a few years.  
UV photons from the star, on the other hand, can heat and excite 
the knots because they carry far less momentum per unit energy.

Most likely, some relatively low-density high-ionization zones 
exist around knot C, with a geometry that somehow allows most 
of the photoionizing energy to be reprocessed in regions slightly 
farther from the star than is the center of the knot.   It would 
be useful to have spatial measurements along a slit perpendicular 
to the orientation that we used, but unfortunately no such data 
exist.

   \begin{figure}  
      \epsscale{0.7}  
   \plotone{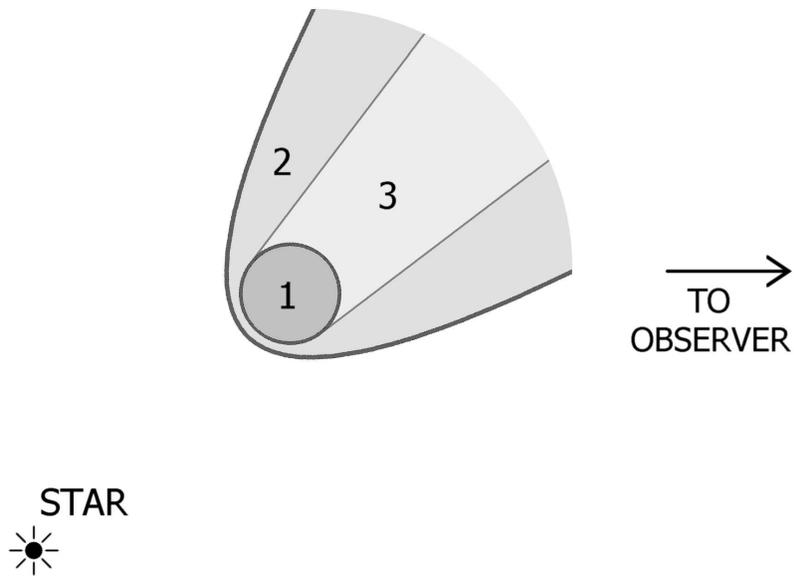} 
   \caption{``Ablation'' caused by photoionization.  Zone 1 is a dense 
      low-ionization cloud,  zone 2 contains outward-flowing gas photoionized 
      by the star,  and zone 3 is shadowed.   Emission lines from zone 2 might 
      on average originate higher than zone 1  in this diagram, see text.  
      This is merely a qualitative sketch of one possibility. } 
   \label{fig:ablateblob}   
   \end{figure}

Figure \ref{fig:ablateblob} shows one specific possibility.    
When a gas condensation is illuminated by 
ionizing photons,  heated material can ``evaporate'' through 
an expansion front.  In favorable circumstances the escaping 
gas flows outward around the condensation, accelerated by 
radiative forces exerted via photoionization:  zone 2 in 
the figure.  One might call this process photo-ablation or 
photo-evaporation, and it can produce a rocket effect that we 
are not concerned with here.  Because zone 2 is less dense 
and is directly exposed to the central star, it should 
produce mainly high-ionization category-C emission lines.
As Figure \ref{fig:ablateblob} illustrates, emission lines from 
zone 2 may appear preferentially on the {\it outer\/} side of 
the condensation itself (zone 1), in our projected view.

Admittedly the densest emission region should be near the expansion 
front, adjoining the inner edge of zone 1.  The basic idea 
nevertheless remains viable, because those inward-facing regions 
may be partially obscured by dust in the condensation, and also 
for reasons involving the density-dependent [\ion{Ne}{3}] and 
[\ion{Ar}{3}] emissivities.  A detailed model of the accelerated 
flow is needed in order to say whether the high-ionization lines 
should exhibit conspicuous velocity differences. If an explanation  
of this type is true, then it is potentially useful because the 
partially observable structure depends on parameters of the knot.

Conceivably a separate, unrelated {\it high-ionization\/} knot lies along 
a line of sight slightly farther from the star than the center of knot C, 
thereby perturbing our [\ion{Ne}{3}] and [\ion{Ar}{3}] measurements.  
But this idea is somewhat artificial,  because 
the hypothetical condensation must have a particular column density 
and scale size in order to produce enough localized 
high-ionization emission without comparable low-ionization 
features.   A large, low-density cloud, for instance, could produce 
[\ion{Ne}{3}] but its spatial gradients would be insufficient for 
the proposed effect.  High-ionization knots 
have not been obvious in the large volume of existing 
STIS data near $\eta$ Car.    If the true explanation is in 
this vein, then it strongly suggests a widespread, highly inhomogeous 
configuration of low-speed ejecta.  One good test would be to examine 
Weigelt knot D, see a later comment below.

In an unconventional view, the Weigelt brightness peaks may 
represent minima in the intervening extinction, not physical 
condensations.   We do not advocate this idea here, and we have 
not attempted to construct such a model, but it has not been 
ruled out and it would fundamentally alter the meanings of Figure 
\ref{fig:zvals} and Table \ref{tab:zlist}.  This question needs 
more comparisons between {\it HST\/} and IR maps (cf.\ \citealt{arti11}).   
Precise Doppler velocities are worth investigating as noted 
in {\S}\ref{ssec:other}.

The spatial profiles in Figures \ref{fig:netprofs1} and \ref{fig:netprofs2}
have other interesting implications.  For example the Weigelt knots 
originally appeared to be small condensations \citep{weig86,hofm88}, but 
in fact they have widths almost half as large as their distances from the 
star ({\S}\ref{ssec:other} above, and Fig.\ 4 in \citealt{arti11}).  
This fact obviously affects theories of their origin and makes their  
proper motions more difficult to quantify ({\S}\ref{sec:method} above).

In Table \ref{tab:zlist}, entries I1 and I2 are two estimates of 
$\hat{z}$ for dust in Knot C, based on published measurements of 
non-spectroscopic 
images.  Various emission lines also contribute, but the images were 
most likely dominated by hot-dust emission for I1 and dust reflection 
for I2.  \cite{ches05} and \citet{arti11}, discussing the near-IR data 
used for position I1, carefully referred to knot C$^\prime$, 
not C, to distinguish between the locations of dust emission and 
reflection.  Thus it makes sense that $\hat{z}$ for knot 
C$^\prime$ (I1 in Table \ref{tab:zlist}) is larger than the value 
for low-excitation emission lines.   The I2 value, on the other 
hand, appears surprisingly large at 237 mas.  It was based on 
{\it HST\/} images \citep{smit04}, and in a simple model it should have 
been in the neighborhood of 220 mas.  We cannot explore this 
question here, but a likely guess is that 
irregularities in the p.s.f.\ and other subtleties in the central 
star image led to systematic errors as hinted 
in {\S}\ref{sec:method} above.  This issue matters because the 
often-quoted \citet{smit04} analysis of the knots' proper motions 
and age relied on the same images and measurement techniques.

The other well-defined Weigelt knot, knot D, should be examined 
for ionization structure in the same way as knot C.  
Unfortunately we have not been able to obtain {\it HST\/} 
observing time for suitably dithered observations.  As a conceivable 
alternative approach, one might employ a large set of existing data.   
Many STIS observations of $\eta$ Car in 1998--2011 
used a slit position angle near 331{\degree} which samples 
knot D, see \citet{mehn10} and refs.\ therein.  At a subpixel 
level, those data have more or less random placements of the star 
and knot along the slit -- a form of accidental spatial dithering. 
However, since  the star/knot brightness ratio has varied and we have no 
assurance that the knot's spatial profile has remained constant,  
a careful investigation along these lines would require more effort 
than the knot C analysis presented above.

In approximately the same manner as our ionization-zone analysis,
{\it HST\/}/STIS data may be useful for two obvious additional 
tasks.    First, this appears to be the best way to measure the 
Weigelt knots' proper motions and age at UV to red wavelengths. 
The main obstacle is a limited temporal baseline, beginning 
no earlier than 1998.  Existing data near slit position angle 
331{\degree} (see above) are pertinent in this connection.  In 
principle, ``virtual dithering'' for each observation date might 
be obtained by using separate 
emission lines which peak at different fractional-row locations 
because of the slope of the spectral trace ({\S}\ref{ssec:trace}).

A second interesting goal would be to map the region within 1{\arcsec} 
of $\eta$ Car in somewhat the same way as \citet{mehn10} 
but with higher spatial resolution.  The fainter knot beyond 
C (Figures \ref{fig:netprofs1} and \ref{fig:netprofs2}) is one 
example of why this would be useful;  multiple brightness 
peaks in the region are very poorly known. This goal requires 
STIS observations that are spatially dithered along the slit,  
arguably more essential than dithering perpendicular to it.    
Unfortunately no such data exist at present, except those 
used in this paper.   

   \vspace{3mm}  

{\it Acknowledgements} \\  
GNR is supported by a Hertz Foundation Fellowship.  
We thank Beth Perriello and other consultants at STScI for 
assistance with the {\it HST\/} observing techniques. 
%%%

                    %%%  ===-===  %%%  
                    %%%  TABLES   %%%

\newpage 
\begin{deluxetable}{rlcccc}   %%%  TABLE 1
   \tabletypesize{\scriptsize}  
\tablecaption{Narrow Emission Lines Measured in Weigelt Knot C
   \label{tab:linelist} }    
\tablewidth{0pt} 
\tablehead{  
  \colhead{Label \&}  &   
  \colhead{ID\tablenotemark{b}}  &  \colhead{$\lambda_\mathrm{vac}$}  & 
  \colhead{Transition\tablenotemark{b}}  &
  \colhead{$E_2$\tablenotemark{c}}  &  \colhead{$E_1$\tablenotemark{c}}     
      \\  
  \colhead{category\tablenotemark{a}}         &  
  \colhead{}        &  \colhead{(\AA)}  & 
  \colhead{}  &  \colhead{(eV)}  &  \colhead{(eV)}  }  %% end of tablehead
\startdata
\  
   1 \ \ D  \ \ &  Fe II               &  2507.55         & 
       $\mathrm{c^{4}F_{7/2}-5p^{6} F_{9/2}}$          & 
       11.167  &   6.222  \\
   2 \ \ D  \ \ &  Fe II               &  2509.10         & 
       $\mathrm{c^{4}F_{7/2}-4p^{4} G_{9/2}}$          & 
       11.164  &   6.222  \\
                          \\  %% spacer line 
   3 \ \ C  \ \ &  {[}Ne III{]} (1F)   &  3869.85         & 
       $\mathrm{2p^{4}\,^{3}P_{2}-2p^{4}\,^{1}D_{2}}$  & 
        3.204  &    0     \\ 
   4 \ \ A  \ \ &  Fe II (3)          &  3939.41         &  
     $\mathrm{a^{4}P_{5/2}-z^{6}D_{5/2}}$              &   
        4.818  &  1.671   \\  
   5 \ \ A  \ \ & {[}Ni II{]} (3F)    &  3994.19         & 
     $\mathrm{a^{2}D_{5/2}-b^{2}D_{5/2}}$              &   
        3.104  &    0     \\    
   6 \ \ C  \ \ &  He I (18)         &  4027.33         & 
     $\mathrm{2p\,^{3}P-5d\,^{3}D}$                    & 
       24.043  &  20.964  \\ 
   7 \ \ A  \ \ & {[}S II{]} (1F)     &  4069.75         & 
    $\mathrm{3p^{3}\,^{4}S_3/2-3p^{3}\,^{2}P_3/2}$     & 
        3.046  &    0     \\    
                          \\  %% extra spacer 
   8 \ \ A  \ \ &  Fe II  (38)         &  4585.12         & 
     $\mathrm{b^{4}F_{9/2}-z^{4}D_{7/2}}$              &  
        5.511  &   2.807  \\  
   9 \ \ A  \ \ &  Fe II  (37)         &  4630.64         & 
     $\mathrm{b^{4}F_{9/2}-z^{4}F_{9/2}}$              & 
        5.484  &   2.807  \\  
  10 \ \ A  \ \ &  {[}Fe II{]} (4F)   &  4640.97         & 
     $\mathrm{a^{6}D_{3/2}-b^{4}P_{1/2}}$              & 
        2.778  &   0.107  \\  
  11 \ \ B  \ \ &  {[}Fe  III{]} (3F)   &  4659.35         & 
     $\mathrm{3d^{6}\,^{5}D_{4}-3d^{6}\,^{3}F2_{4}}$   & 
        2.661  &    0     \\  
  12 \ \ B  \ \ &  {[}Fe III{]} (3F)   &  4702.85         & 
     $\mathrm{3d^{6}\,^{5}D_{3}-3d^{6}\,^{3}F2_{3}}$   & 
        2.690  &   0.054  \\  
  13 \ \ A  \ \ &  {[}Fe II{]} (4F)   &  4729.39         & 
     $\mathrm{a^{6}D_{5/2}-b^{4}P_{3/2}}$              & 
        2.704  &   0.083  \\  
  14 \ \ A  \ \ &  {[}Fe II{]} (20F)  &  4776.05         & 
     $\mathrm{a^{4}F_{9/2}-b^{4}F_{7/2}}$              & 
        2.828  &   0.232  \\  
  15 \ \ A  \ \ &  {[}Fe II{]} (20F)  &  4815.88         & 
     $\mathrm{a^{4}F_{9/2}-b^{4}F_{9/2}}$              & 
        2.807  &   0.232  \\  
                          \\  %% spacer line 
  16 \ \ C  \ \ &  He I  (10)         &  7067.20         & 
     $\mathrm{2p\,^{3}P-3s\,^{3}S}$                    & 
       22.718  &  20.964  \\ 
  17 \ \ C  \ \ &  {[}Ar III{]} (1F)   &  7137.76         & 
     $\mathrm{3p^{4}\,^{3}P_{2}-3p^{4}\,^{1}D_{2}}$    & 
        1.737  &    0     \\  
  18 \ \ A  \ \ &  {[}Fe II{]} (14F)  &  7157.13         & 
     $\mathrm{a^{4}F_{9/2}-a^{2}G_{9/2}}$              & 
        1.964  &   0.232  \\ 
  19 \ \ A  \ \ &  {[}Fe II{]} (14F)  &  7173.98         & 
     $\mathrm{a^{4}F_{7/2}-a^{2}G_{7/2}}$              & 
        2.030  &   0.301  \\     
\enddata  

  \tablenotetext{a}{A,B,C = low, moderate, or high ionization, D = ``exotic,''  
        see text. }    
  \tablenotetext{b}{\citet{zeth01,zeth12}.} 
  \tablenotetext{c}{\url{http://physics.nist.gov/PhysRefData/ASD/levels\_form.html}.}

\end{deluxetable}

\newpage

                       %%%  ===-===  %%%  

\begin{deluxetable}{cccrc}   %%%  TABLE 2
    \tabletypesize{\scriptsize}  
\tablecaption{ Dithered HST/STIS/CCD Observations\tablenotemark{a} 
    \label{tab:obslist} }  
\tablewidth{0pt} 
\tablehead{  
  \colhead{Root}     &   \colhead{Wavelength}   &  Position\tablenotemark{c}   
  &  ${\Delta}t$\tablenotemark{d}   &  $N$\tablenotemark{e}   
      \\    
  \colhead{ number\tablenotemark{b}}   &   \colhead{Range (\AA)}  &  
  \colhead{(pixels)}   &  \colhead{(s)}  &       
  }   %% end of tablehead
\startdata
  ob6064150   &  2480--2680  &  510.129  & 200   &  2  \\    
  ob6064160   &     $''$     &  514.641  & 200   &  2  \\   
              &              &           &       &     \\   
  ob60640s0   &  3795--4075  &  514.995  &   6   &  2  \\   
  ob60640u0   &     $''$     &  519.488  &   6   &  2  \\    
  ob60640t0   &     $''$     &  515.031  &  54   &  3  \\   
  ob60640v0   &     $''$     &  519.534  &  54   &  3  \\   
              &              &           &       &     \\   
  ob60640j0   &  4565--4845  &  513.161  &  10   &  2  \\     
  ob60640l0   &     $''$     &  517.648  &  10   &  2  \\  
  ob60640k0   &     $''$     &  513.183  &  60   &  4  \\   
  ob60640m0   &     $''$     &  517.681  &  60   &  4  \\  
              &              &           &       &     \\  
  ob60641b0   &  7000--7565  &  517.615  &  21   &  3  \\       
  ob60641d0   &     $''$     &  522.122  &  21   &  3  \\   
\enddata  
  \tablenotetext{a}{Obs.\ 2010 March 3,  
  MJD 55258.7 = J2010.17, HST Program GO 11612 
  (PI = K.\ Davidson). } 
  \tablenotetext{b}{ID label in the HST archive.} 
  \tablenotetext{c}{``Position'' means the measured CCD row number of  
     the central star continuum ($x_A$ in Fig.\ \ref{fig:method} 
     and {\S}\ref{ssec:trace})  
     at column 494 near the middle of the CCD.  The nominal 
     row width was 50.71 mas, the slit width was 100 mas, its position 
     angle was 301.95{\degree}, and an initial peakup was used to place 
     the central star as close to the slit midline as possible.}  
  \tablenotetext{d}{Integration time (= total exposure time).}   
  \tablenotetext{e}{$N$ = CR-SPLIT, the number of separate 
     exposures combined to make an ``observation.''}   
\end{deluxetable}

                    %%%%  ===-===                   

\newpage 
\begin{deluxetable}{rccccccc}   %%%  TABLE 3
    \tabletypesize{\scriptsize}  
\tablecaption{Measured Positions \label{tab:measures} }    
\tablewidth{0pt} 
\tablehead{  
   \multicolumn{3}{c}{Feature and category\tablenotemark{a}}  &         
            \colhead{$u$\tablenotemark{b}}  &  Peak\tablenotemark{c}    &  
            $z_c$\tablenotemark{d}   &  $z_m$\tablenotemark{d}  
            & err\tablenotemark{e}              
     }  %% end of tablehead  
\startdata
  1 & Fe II        $\lambda$2508  & D &  183  &  4350    
      &  4.426   &  4.394  &  $\pm$.06:     \\ 
  2 & Fe II        $\lambda$2509  & D &  193  &  5660  
      &  4.256   &  4.290  &  $\pm$.06:     \\
            \\  %%% spacer line                 
  3 & {[}Ne III{]}  $\lambda$3870  & C &  269  &  3320  
      &  4.844   &  4.740  &  $\pm$.055      \\
  4 & Fe II        $\lambda$3939  & A &  519  &   890  
      &  3.939   &  3.990  &  $\pm$.100      \\  
  5 & {[}Ni II{]}  $\lambda$3994  & A &  717  &   670  
      &  4.270   &  4.269  &  $\pm$.120       \\  
  6 & He I        $\lambda$4027  & C &  836  &   730   
      &  4.258   &  4.332 &   $\pm$.120      \\ 
  7 & {[}S II{]}   $\lambda$4070  & A &  990  &  1310  
      &  4.138   &  4.216 &   $\pm$.085      \\  
            \\  %%% spacer line   
  8 & Fe II         $\lambda$4585  & A &   91  &  2570   
      &  3.869   &  3.946 &   $\pm$.060      \\  
  9 & Fe II        $\lambda$4631  & A &  255  &  1830   
      &  3.698   &  3.671 &   $\pm$.070      \\  
 10 & {[}Fe II{]}  $\lambda$4641  & A &  293  &   890  
      &  4.073   &  4.130 &   $\pm$.100      \\  
 11 & {[}Fe III{]}  $\lambda$4659  & B &  359  &  1880  
      &  4.182   &  4.310 &   $\pm$.070      \\  
 12 & {[}Fe III{]}  $\lambda$4703  & B &  516  &   770   
      &  4.384   &  4.325 &   $\pm$.110      \\  
 13 & {[}Fe II{]}  $\lambda$4729  & A &  612  &  1860  
      &  3.895   &  3.997 &   $\pm$.070      \\  
 14 & {[}Fe II{]}  $\lambda$4776  & A &  781  &   850  
      &  3.766   &  3.721 &   $\pm$.110      \\  
 15 & {[}Fe II{]}  $\lambda$4816  & A &  926  &  2640  
      &  3.990   &  4.026 &   $\pm$.060       \\  
            \\  %% spacer line  
 16 & He I         $\lambda$7067  & C &  127  &  2430  
      &  4.396   &  4.332 &   $\pm$.060      \\   
 17 & {[}Ar III{]}  $\lambda$7138  & C &  254  &   780   
      &  5.102   &  5.023 &   $\pm$.110      \\  
 18 & {[}Fe II{]}  $\lambda$7157  & A &  289  &  3330  
      &  4.152   &  4.161 &   $\pm$.055      \\  
 19 & {[}Fe II{]}  $\lambda$7174  & A &  319  &   970   
      &  4.084   &  4.091 &   $\pm$.100      \\  
\enddata  
  \tablenotetext{a}{See Table \ref{tab:linelist}.}
  \tablenotetext{b}{Approximate STIS/CCD column (1--1024), 
         cf.\ Table \ref{tab:obslist}.} 
  \tablenotetext{c}{Approximate maximum net counts per data point 
         in Knot C.} 
  \tablenotetext{d}{Centroid and midpoint distances from the 
         star, expressed in CCD pixels ($\sim$50.7 mas). }   
  \tablenotetext{e}{Estimated r.m.s.\ statistical errors in $z$, 
         not including systematic effects which may be larger. }  
\end{deluxetable}

                       %%%  ===-===  %%%  

\begin{deluxetable}{cccc}   %%%  TABLE 4
    \tabletypesize{\scriptsize}  
\tablecaption{ Measured Distances from the Star  \label{tab:zlist} }  
\tablewidth{0pt} 
\tablehead{  
  \colhead{Ionization}   &  \colhead{$n_{\,\mathrm{lines}}$\tablenotemark{a}}   
  & \colhead{Median $\hat{z}$}   &  \colhead{Range of}     
     \\  
  \colhead{category\tablenotemark{a}}  &   &  
  \colhead{(mas)\tablenotemark{b}}   &   \colhead{values (mas)}      
  }   %%  end of tablehead 
\startdata
  A                    &  11  &  204  &  187 -- 217  \\   
  B                    &   2  &  218  &  216 -- 221  \\  
%%  C1\tablenotemark{c}  &   4  &  232  &  218 -- 257  \\  
  C  &   4  &  232  &  218 -- 257  \\  
  C\tablenotemark{c}  &   2  &  249  &  243 -- 257  \\  
  D                    &   2  &  220  &  216 -- 224  \\  
           \\  
  I1\tablenotemark{d}  &  (dust?)  &  (229) &  ---  \\  
  I2\tablenotemark{e}  &  (dust?)  &  (237) &  ---    
\enddata  
  \tablenotetext{a}{See Table 1.}
  \tablenotetext{b}{Projected distance from star, assuming 
      CCD pixel-width = 50.71 mas. }   
  \tablenotetext{c}{Omitting He I for reasons noted in {\S}4.}   
  \tablenotetext{d}{I1 = location of knot C$^\prime$ in Chesneau's 
     2002--2005 near-IR image data, corrected for expansion to 2010.  
     See Fig.\ 4 in \citet{arti11}. }
  \tablenotetext{e}{I2 = location in 2010 ``predicted'' from HST image 
     data, probably less accurate than I1.  See Fig.\ 11 in 
     \citet{smit04}. } 
\end{deluxetable}

                    %%%%  ===-=== FIGURES                  
\newpage

  \newpage 

%% \begin{figure}  
%%   %\epsscale{1.2}  
%% \plotone{fig6zvals.eps} 
%% \caption{Measured distances of emission peaks from the star, 
%%   sorted by ionization category.  These are the results listed 
%%   in Table \ref{tab:measures};  filled symbols represent 
%%   higher-quality features.  Circles and squares indicate 
%%   $z_c$ and $z_m$ respectively, see text.}   
%% \label{fig:zvals}   
%% \end{figure}  

 \newpage 

%% \begin{figure}  
%%   %\epsscale{1.2}  
%% \plotone{fig7ablation.eps} 
%% \caption{``Ablation'' caused by photoionization.  Zone 1 is a dense 
%%    low-ionization cloud,  zone 2 contains outward-flowing gas photoionized 
%%    by the star,  and zone 3 is shadowed.   Emission lines from zone 2 might 
%%    on average originate higher than zone 1  in this diagram, see text.  
%%    This is merely a qualitative sketch of one possibility. } 
%% \label{fig:ablateblob}   
%% \end{figure}  

               %%%  ===-===  %%% 

\end{document}